%% file: Forsterite_Xtal_160GPa.tex
\documentclass[11pt, letterpaper, twocolumn]{article} 

\input{structure.tex} 

\title{Crystalline forsterite to 160 GPa:  the striking metastability of one of Universe's most abundant minerals}

\author{
	\authorstyle{Barbara Lavina\textsuperscript{1,\S},  Minta C. Akin\textsuperscript{2}, Yue Meng\textsuperscript{3,\dag}, Vitali Prakapenka\textsuperscript{1}}
	\newline\newline 
	\textsuperscript{1}\institution{Center for Advanced Radiation Sources, The University of Chicago, Chicago, IL, USA}\\ 
	\textsuperscript{2}\institution{Lawrence Livermore National Laboratory, Livermore, CA, USA}\\ 
	\textsuperscript{3}\institution{X-ray Science Division, Advanced Photon Source, Argonne National Laboratory, Lemont, IL, USA}\\ 
	\textsuperscript{\S}\institution{corresponding author: lavina@uchicago.edu; https://orcid.org/0000-0002-8556-7916}\\
	\textsuperscript{\dag}\institution{current address: mengyue.cc@gmail.com}
}
\date{} 

\begin{document}

\maketitle 

\thispagestyle{firstpage} 


\lettrineabstract{Among Universe's most consequential events  are large impacts generating rapidly-evolving extreme pressures and temperatures. 
Crystalline and amorphous forms of (Mg, Fe)$_2$SiO$_4$ are abundant and widespread, within planets and in space. The behavior of these minerals is expected to deviate from thermodynamic equilibrium in many of the processes that are critical  to the formation and evolution of planets, particularly shock events. 
It has been shown that ambient-temperature static compression and shock compression behavior of forsterite (the Mg end-member phase stable at ambient conditions) follow similar deformation pathways up to about 90 GPa. 
To further the  understanding of the behavior of the silicate under extreme conditions, we statically compressed a crystal of forsterite up to 160.5 GPa, far beyond the compound's  stability field, and probed its long-range ordering  with synchrotron microdiffraction. 
We found that forsterite  retains long-range ordering up to the highest pressure reached. Forsterite III, emerging at about 58 GPa \cite{Finkelstein:2014}, persists in compression to 160.5 GPa and  in decompression down to about 13 GPa, for a rare combined occurrence of a metastable phase of nearly 150 GPa. These observations dispute earlier reports of pressure-induced amorphization and are a unique testimony of the resilience of the crystalline state in quasi hydrostatic compression.
We confirm that highly disordered forsterite can  be obtained from the decompression of forsterite III as suggested from the substantial loss of long-range ordering observed at 7 GPa after further decompression. Such kinetic pathway may explain how synthetic olivine glass  have been obtained in shock experiments  and could be a mechanism of generation of  amorphous forsterite in cosmic dust. The 120 GPa  Hugoniot discontinuity  finds no correspondence in our data, marking a departure from the parallelism between  static "cold compression" and dynamic compression. 
}\\

\begin{figure} 
\centering
\includegraphics[width=1.0\linewidth]{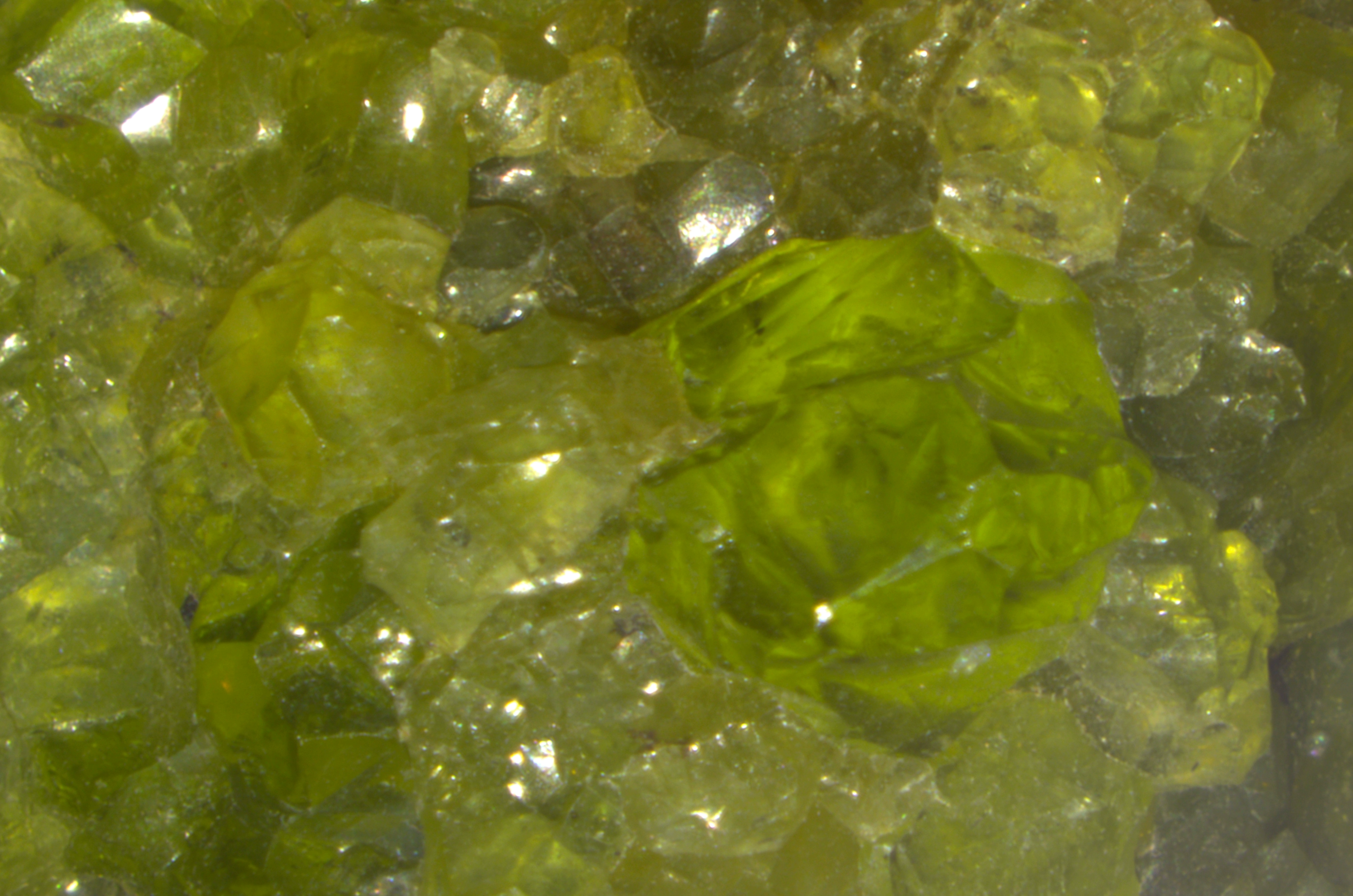}
\caption{Aggregate of S. Carlos olivine crystals.}
\label{photo}
\end{figure}

\lettrine{O}livine, (Mg, Fe)$_2$SiO$_4$  (Figure \ref{photo}), is abundant and widespread in Earth, terrestrial planets, meteorites and cosmic dust. It is a major constituent of rocky planets' mantles, its polymorphism and breakdown  causing  major Deep-Earth discontinuities \cite{Duffy:2015, Irifune:2015}. The mineral is abundant in meteorites \cite{Weisberg:2006, Mittlefehldt:2008}, comets \cite{Hanner:2004}  and the interstellar dust \cite{Bradley:2003aa,  Henning:2010aa}. 
Upon compression, after transforming into wadsleyite ($\beta$ beta-Mg$_2$SiO$_4$) and then into ringwoodite ($\gamma$-Mg$_2$SiO$_4$), the silicate becomes metastable with respect to breakdown into MgO + MgSiO$_3$ around 23 GPa. While  olivine polymorphs may survive greater pressures within cold subduction slabs \cite{KIRBY:1991aa, Kaneshima:2007aa, Zhang:2019aa}, they are unlikely to be preserved in bulk for extended time beyond their stability pressure in planetary interiors. 
Metastable forms of (Mg, Fe)$_2$SiO$_4$ persist at pressures much greater than the compound’s stability region because of the high energy barrier of the dissociation reaction. In nature,  large amounts of  short-lived over-compressed metastable  (Mg, Fe)$_2$SiO$_4$ presumably occur during  meteoritic and planetesimal impacts, events essential to the formation, evolution, and chemical differentiation of Earth and other planets. Metastable phases produced during shock compression can be preserved in rapidly cooled materials, such as impact ejecta. Thus, to model planetary-forming impacts  and decipher  the petrology of shock metamorphosed materials, it is  important to understand the behavior of olivine far beyond its thermodynamic stability. 
Metastable phases of  (Mg, Fe)$_2$SiO$_4$ include amorphous and crystalline forms. Amorphous forsterite is the most abundant silicate  in the interstellar dust \cite{Bradley:2003aa,  Henning:2010aa}. Synthetic olivine glasses were first reported in recovered shock-compressed materials \cite{JEANLOZ:1977aa}. The observation  triggered numerous efforts to understand the factors controlling defects and amorphization processes aimed at deciphering the geological record of naturally-occurring shock-compressed materials \cite{Jeanloz:1980aa, Syono:1981, Langenhorst:1999, Farrell-Turner:2005, Gillet:2007, SCHMITT:2000, Walton:2013, Wang:2017, Takenouchi:2018aa}. Amorphous olivine can also be produced in static compression experiments.  A strong composition dependence of pressure-induced amorphization (PIA) under non-hydrostatic compression  was suggested by Andrault et al. \cite{Andrault1995}; the PIA of forsterite  was observed at about 50 GPa,  soon after an unidentified phase transition, in agreement  with earlier TEM analysis of recovered samples \cite{Guyot:1992aa}.
The occurrence of metastable crystalline forms of forsterite was recently extended to 90 GPa in ambient-temperature static compression  \cite{Finkelstein:2014}  and to 73 GPa and about 3000 K in dynamic compression  \cite{Newman:2018aa}. Finkelstein et al.\cite{Finkelstein:2014} revealed the occurrence of two first-order structural phase transitions and the persistence of a crystalline state up to about 90 GPa. In particular, at about 50 GPa forsterite transformed into a triclinic phase, forsterite II, followed by orthorhombic forsterite III above 58 GPa. Calculations suggest that the polymorph, while remaining unstable with respect to the breakdown phases, becomes energetically favored compared to other known polymorphs around 25 GPa \cite{Zhang:2019aa}. The first \emph{in-situ} x-ray diffraction data collected on dynamically compressed forsterite \cite{Newman:2018aa} demonstrated that forsterite I persists up to at least 44 GPa.  Not only  chemical segregation but also  polymorphism is kinetically hindered along the forsterite Hugoniot \cite{Mosenfelder:2007aa, Gillet:2013}; furthermore, forsterite III was found at ~73 GPa and 3000 K, resolving a longstanding debate on whether the compound segregates at such conditions \cite{Jeanloz:1980aa, Syono:1981}.

The seemingly scattered and conflicting observations summarized above  reflect the complexity of factors affecting the metastability of olivine, including stress, temperature, strain rate,  composition.  Natural shock events cannot be entirely replicated; the timescale of natural and laboratory shock events, for instance, differs by orders of magnitude, and the shock conditions within natural objects (planets, planetesimals, meteorites, etc) are extremely heterogeneous \cite{Xie:2001aa, Sharp:2006aa}. While such complexity makes it difficult to model the mineral’s behavior, it also means that olivine can be  a  marker of shock history and, given its abundance, a mineral affecting the shock itself.  It is therefore important to achieve a fundamental understanding of the behavior of olivine  under pressure, including its metastability. With this in mind, we performed a quasi-hydrostatic compression study of pure forsterite to 160.5 GPa, to   further explore its "cold compression" polymorphism and learn whether the material eventually becomes amorphous. Static compression observations can facilitate the interpretation of the discontinuities in the Hugoniot as in the case of forsterite III \cite{Newman:2018aa, Finkelstein:2014}, and while the dynamic behavior of forsterite is being explored under conditions of super-Earth  interiors \cite{Sekine:2016aa, Bolis:2016aa, Root:2018aa}, the nature of the discontinuity at about 120 GPa is still debatable as in-situ structural observations have not yet been made. 
\\       

\begin{figure*}[th]
\includegraphics[width=1.0\linewidth]{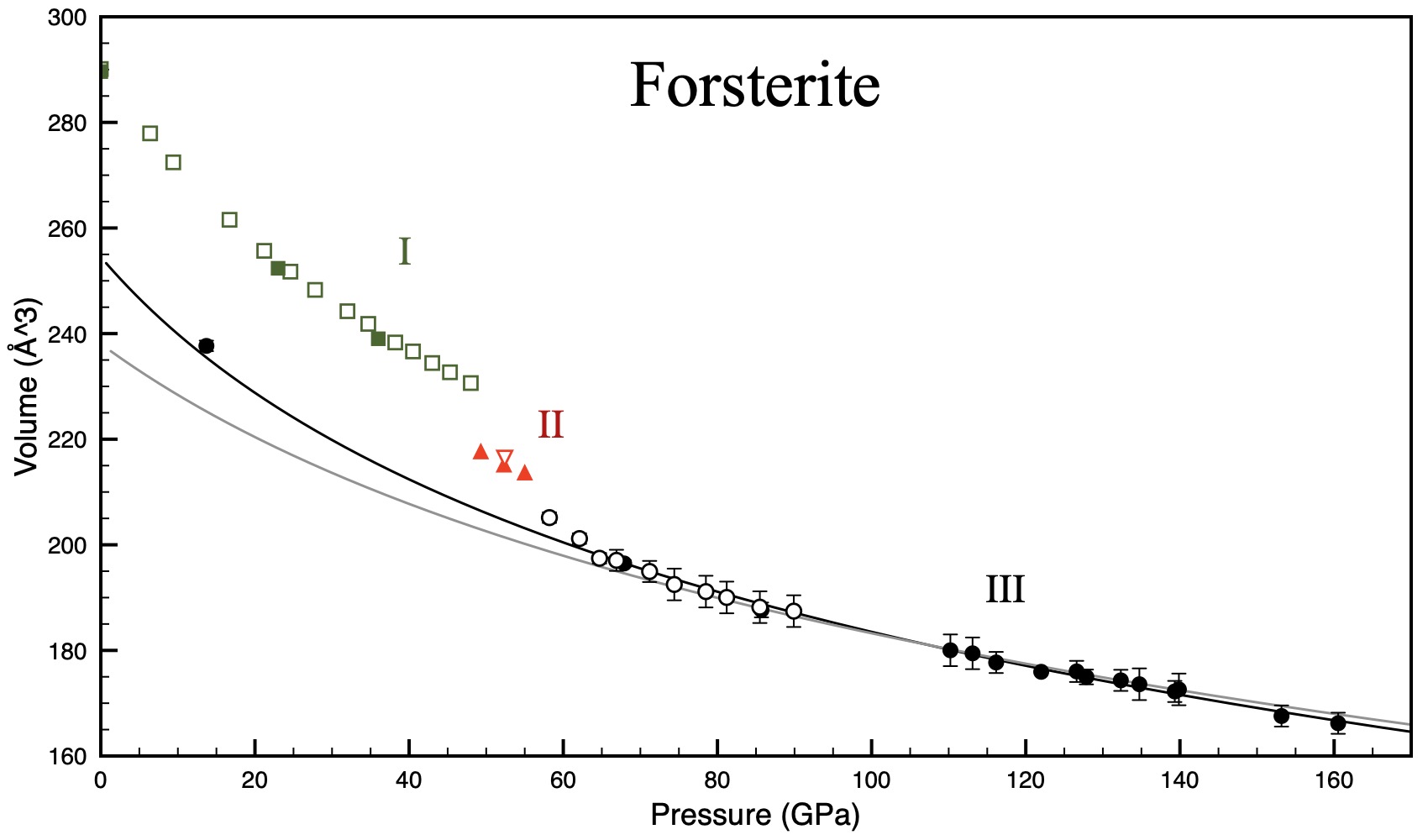}
\caption{ Unit-cell volume of three forsterite polymorphs as a function of pressure, forsterite I: squares, forsterite II: triangles, forsterite III: circles. Open symbols show literature data \cite{Finkelstein:2014}, solid symbols represent this work's data. The black line shows this work  EoS, the grey line shows a calculated EoS from the literature \cite{Zhang:2019aa}.}
\label{PVall}
\end{figure*}

We performed few measurements  in the previously-explored pressure range \cite{Finkelstein:2014} for consistency. The diffraction patterns collected at ambient conditions show sharp Bragg peaks. Calculated lattice parameters are in good agreement with the literature (Figure \ref{FoIabc}) showing that the sample is  suitable for the study. Lattice parameters of forsterite I were also measured at 23 and 36 GPa and a  structural refinement was performed at 23 GPa (Table \ref{tab}). Between 49.3 and 55.0 GPa, we observed a twinned triclinic crystal with rather sharp Bragg peaks, consistent with the description of forsterite II by Finkelstein et al. \cite{Finkelstein:2014}. The following phase transition to forsterite III is associated with   peak broadening. Because of the limited number of observed reflections and stronger parasitic scattering in our experiment, we could not attempt a reassessment or refinement of the crystal structure of the high-pressure polymorph; our data is nonetheless consistent with the orthorhombic \emph{C} lattice proposed by Finkelstein et al. \cite{Finkelstein:2014}. The very satisfactory agreement with the literature  of all measurements for all phases, including lattice types, pressure dependence of lattice parameters, (Figure \ref{FoIabc}, \ref{FoIIIabc}) and transition mechanisms, namely twinning in forsterite II and peak broadening in forsterite III, show that the experiments are highly reproducible, implying that all crystals experienced similar stress conditions, and none of the specimens bridged the anvils below 90 GPa. The continuous behavior of lattice  parameters throughout the forsterite III range (Figure~\ref{FoIIIabc})  indicates that anvils bridging most likely did not occur above 90 GPa either. The compression of the forsterite III  above 90 GPa caused further continuous broadening of Bragg peaks but no changes in patterns that could indicate further structural phase transitions up to the highest pressure reached, 160.5 GPa. The peak broadening is most likely due to the defects accumulated during the first-order phase transitions and the stresses between the crystal's mosaic domains rather than the modest medium stiffening \cite{Dewaele:2007aa}. We conclude that pure forsterite retains a crystalline state in static quasi-hydrostatic compression up to such high pressures, and that it remains in the forsterite III phase.\\
\begin{figure*}[th]
\centering
\includegraphics[width=1.0\linewidth]{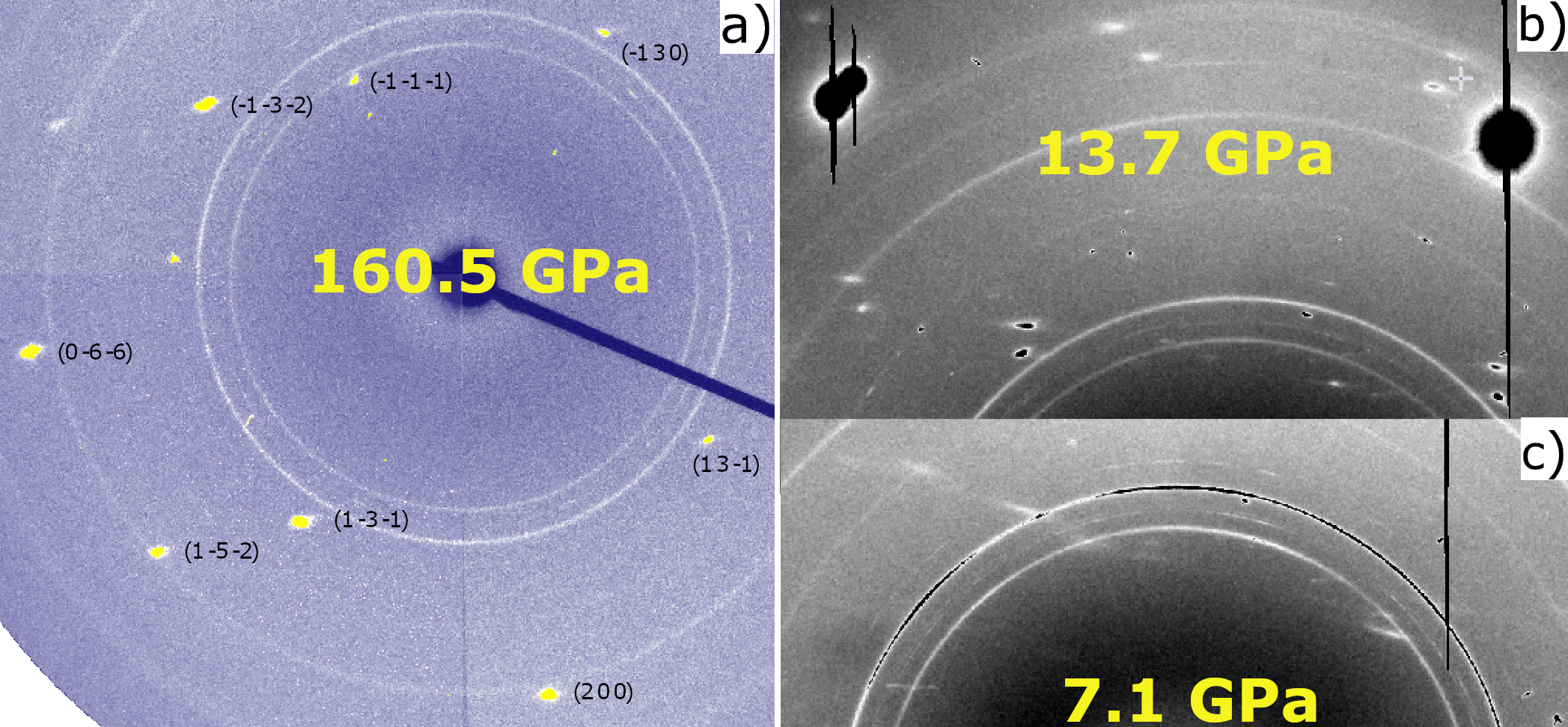}
\caption{Examples of diffraction images collected at different pressures. a) Diffraction pattern of forsterite III at 160.5 GPa; b) Forsterite III collected at 13.7 GPa after decompression; c) Pattern collected at 7.1 GPa upon  decompression.} 
\label{diffraction}
\end{figure*}

Upon decompression, Finkelstein et al. \cite{Finkelstein:2014} observed  strong hysteresis of the phase transitions and eventually amorphization at ambient conditions. We compressed a crystal to 67.9 GPa, obtaining forsterite III, and collected two data-sets in decompression. We  observed forsterite III at 13.7 GPa (Figures \ref{PVall}, \ref{FoIIIabc}). Thus, combining compression and decompression, we discovered that the metastable phase,  the result of two first-order phase transitions, shows a remarkable persistence of nearly 150 GPa,  remaining a  single-crystal throughout the journey. Bragg peaks are replaced by broad, discontinuous diffuse scattering rods roughly aligned with the preexisting forsterite III reciprocal lattice rows   upon further pressure release to 7.1 GPa (Figure \ref{diffraction}). The sample is not fully amorphous but experienced a considerable loss of long-range ordering. Further investigations of recovered materials, including TEM analysis, could help understand whether the defect structure of  materials recovered from quasi-hydrostatic, non-hydrostatic, and dynamic  compression shows distinctive microstructural features that could be used to decipher the geological history of natural specimens.
An accurate determination of the forsterite III  equation of state (EoS) is beyond the scope of this study, nonetheless,  extending Forsterite III pressure range compressibility data allowed us to re-evaluate a 3$^{rd}$ order Birch-Murnaghan EoS. We estimate the isothermal bulk modulus $K_{0}$=147   (6) GPa when $V_{0}$ =256 (2) \AA$^{3}$  and $K_{0}$' =4.3 (Figure \ref{EoS}). These values ought to be considered a preliminary assessment of the EoS of forsterite III, since fitting parameters are poorly constrained given the scarcity of data at low pressure.\\

Further experiments to examine the effects of elevated temperatures on forsterite stability would be valuable in interpreting shock data and in constructing an improved forsterite equation of state relevant to planetary mantles, as would experiments to study dissociation and amorphization during decompression. Likewise, future studies and simulations on dissociation mechanism, shearing and defect accumulation during phase transitions, and the effects of strain/defect accumulation upon peak broadening would be valuable in improving experimental data interpretation and in better understanding processes occurring in materials at extreme conditions. Finally, determining the effect of iron and water contents on the material response is necessary to understand the behavior of geological samples.\\

Crystalline forsterite was observed in this study at pressure about  two times higher  than early reports of amorphization in non-hydrostatic conditions \cite{Guyot:1992aa, Andrault1995} and about three times greater than the PIA of San Carlos olivine observed in quasi-hydrostatic conditions \cite{Santamaria-Perez:2016aa}.  While stress anisotropy and composition are known to strongly affect amorphization, the extent of such influence documented in this study is striking. It should be noted  that the detection of PIA  might  be biased by experimental details and probing techniques \cite{Machon:2014aa}.  Indirect evidence of amorphization, namely the absence of crystalline features in  experimental data rather than direct observation of amorphous states, is particularly prone to controversy.   Our  determination of a crystalline state to pressures as high as 160.5 GPa is robust.  To the best of our knowledge, our report is a unique example of a metastable  crystal, obtained after two first-order phase transitions,  withstanding a Mbar of compression (and possibly more) and about half a Mbar of decompression, for a combined observed  occurrence of about 1.5 Mbar. We  observed a major loss of long-range ordering at 7.1 GPa, evidenced by the appearance of  diffuse scattering features. In forsterite, the decompression of the high-pressure polymorph is evidently  more effective than compression in frustrating long-range ordering. These rare observations have general implications on PIA, metastability, the effect of first-order transitions on crystals and material design.\\

The forsterite Hugoniot is characterized by several discontinuities,  not all  understood.  The “mixed region” occurring between $\sim$50 and 120 GPa \cite{Syono:1981, Luo:2004aa, Mosenfelder:2007aa} was  shown to include forsterite III \cite{Newman:2018aa}. The \emph{in-situ} diffraction measurements also showed that ~25\%  of forsterite III might be already present at 44 GPa, which would be consistent with the theoretically-predicted onset of stabilization at $\sim$25 GPa \cite{Zhang:2019aa}, while at 73 GPa  only forsterite III was detectable.  Up to 73 GPa the Hugoniot series of phase transitions  appears to reflect the static ambient-temperature phase sequence  (though shifted to higher pressures \cite{Gillet:2013}) rather than the equilibrium sequence as earlier suggested. In other words, there appears to be a similarity between  time-hindered transformations in shocked forsterite and temperature-hindered transformations in statically compressed forsterite in such pressure regime. 
As we find no further transitions beyond 120 GPa, we cannot explain the nature of the “high pressure phase” regime beginning at about 120 GPa. It is however reasonable for a transition,  possibly including segregation and incongruent melting, that remains kinetically hindered in static compression to occur in shock compression, given the very high temperatures reached in the Hugoniot at this pressure.\\

If the parallelism between the ambient-temperature static  and dynamic compression  deformation mechanisms were to hold in decompression, namely if the transition to forsterite III were to be irreversible upon shock decay, highly disordered materials would form after forsterite is shocked into the forsterite III regime and rapidly decompressed. Thus, the occurrence of forsterite I  would place an upper limit on the peak pressure experienced by shocked rocks. The pressure-temperature-time paths followed by shocked materials in decompression are however variable and complex. 
Pressure, temperature and strain rates vary greatly between shocks, particularly within heterogeneous materials; indeed multiple polymorphs of the same compound are frequently observed in  meteorites in close proximity \cite{Sharp:2006aa}. A very rapid decompression and cooling of forsterite could hinder its equilibration  and  result in amorphization. In contrast, sustained high temperatures would  allow the material to equilibrate, producing the equilibrium assembly at the corresponding   conditions. Sustained high temperatures at modest pressures would allow the re-crystallization of forsterite I. Evidences are difficult to reconcile with this scenario, as diaplectic olivine has not been identified in meteorites \cite{Langenhorst:2002}, and olivine, though deformed and fractured but not re-crystallized, has been recovered from shock experiments above 73 GPa \cite{Jeanloz:1980aa, Xie:2001aa}. Nonetheless, such kinetic path could explain the formation of patchy diaplectic olivine earlier observed \cite{JEANLOZ:1977aa}. 
Given the abundance of olivine in rocky planets, the non-equilibrium shock behavior of the mineral certainly does influence the dynamic and aftermath of the shock, and should be taken into account when modeling such events.
The abundant interstellar amorphous forsterite is believed to be the result of ion bombardment and radiation damages on nanocrystals condensed in cold stellar atmospheres \cite{Jones}. As extreme conditions are very common in the Universe, processes that include high-pressure polymorphism may also be considered to explain the origin of amorphous forsterite in space.
\subsection*{Methods}
Diffraction experiments were performed on synthetic crystal of forsterite at the 16-ID-B and 13-ID-D beamlines of the Advanced Photon Source, Argonne National Laboratory. The gem-quality samples were purchased from Morion Company. We used wide-opening piston-cylinder diamond anvil cells (DAC) equipped with conical diamonds to generate pressure and rhenium gaskets to confine the sample chamber between the anvils’ opposing tips. The pressure transmitting medium (helium) was loaded using the GSECARS/COMPRES gas-loading system \cite{Rivers:2008p1971}. Ruby spheres were placed in the sample chamber to monitor pressure during gas loading and to monitor load increases up to about 70 GPa \cite{MAO:1986}. The  experiment targeting lower pressures was performed with 250 $\mu$m diameter flat culets diamonds, whereas anvils with beveled culets of 150 $\mu$m inner and 300 $\mu$m outer diameter were used for the experiment that reached the highest pressure in this experiment. Gaskets indentation thickness and hole diameters measured 25 $\mu$m and 150 $\mu$m in the first case and 20 $\mu$m and 80 $\mu$m in the latter. Monochromatic (0.2952, 0.34454 and 0.40663 \AA) microfocused (3 to 7 $\mu$m FWHM) x-ray beams were used to probe the sample under high-pressure. Mar165 CCD detectors, calibrated using CeO$_2$ or LaB$_6$ powder patterns and Fit2D \cite{Hammersley:1997} or Dioptas \cite{Prescher:2015aa} were used to collect diffraction images. The powder patterns of the pressure gauges were processed using GSASII \cite{Toby:2013a}. Data were collected with the rotation method with 1-2$^{\circ}$ step-scan for a total range of 56$^{\circ}$ to 66$^{\circ}$. Single-crystal diffraction data were processed with GSE-ADA, RSV \cite{Dera:2013} and  Crysalis \cite{crys}. Pressure was calculated using the compressibility of gold \cite{Takemura:2008aa} and platinum \cite{Dewaele:2004uq}.

\subsection*{Acknowledgements}
\small We are grateful to Sally Lee for her contribution to data analysis and Silvia Lavina for proofreading. We thank Paul Asimow for providing the forsterite sample. 
The experimental work was conducted at HPCAT (Sector 16), and GSECARS
(Sector 13), Advanced Photon Source (APS), Argonne National Laboratory.
HPCAT operations are supported by DOE-NNSA's Office of Experimental Sciences.
GeoSoilEnviroCARS is supported by the National Science Foundation - Earth
Sciences (EAR-1634415) and Department of Energy- GeoSciences
(DE-FG02-94ER14466).
This research used resources of the Advanced Photon Source, a U.S. Department
of Energy (DOE) Office of Science User Facility operated for the DOE Office of
Science by Argonne National Laboratory under Contract No. DE-AC02-06CH11357.
Use of the COMPRES-GSECARS gas loading system was supported by COMPRES under
NSF Cooperative Agreement EAR-1606856 and by GSECARS through NSF grant
EAR-1634415 and DOE grant DE-FG02-94ER14466. 
This work was partially supported by SEES  (EAR–2223273). 
A portion of this work was performed under the auspices of the U.S. Department of Energy by Lawrence Livermore National Laboratory under Contract DE-AC52-07NA27344  LLNL-JRNL-810112.

\subsection*{Author contributions} \small B.L. designed and performed the experiment, interpreted the data, wrote the paper; M.C.A. conceived the study, interpreted the data and contributed to the paper; Y.M. and V.P. designed and built the experimental setups used in this experiment.
\printbibliography[title={{Bibliography}}]
\appendix
\renewcommand\thefigure{\thesection S\arabic{figure}}    
\renewcommand\thetable{\thesection S\arabic{table}}    

\section*{Supplementary information}
\setcounter{figure}{0}

\begin{figure}[h]
\centering
\subsection*{Forsterite I}
\includegraphics[width=1.0\linewidth]{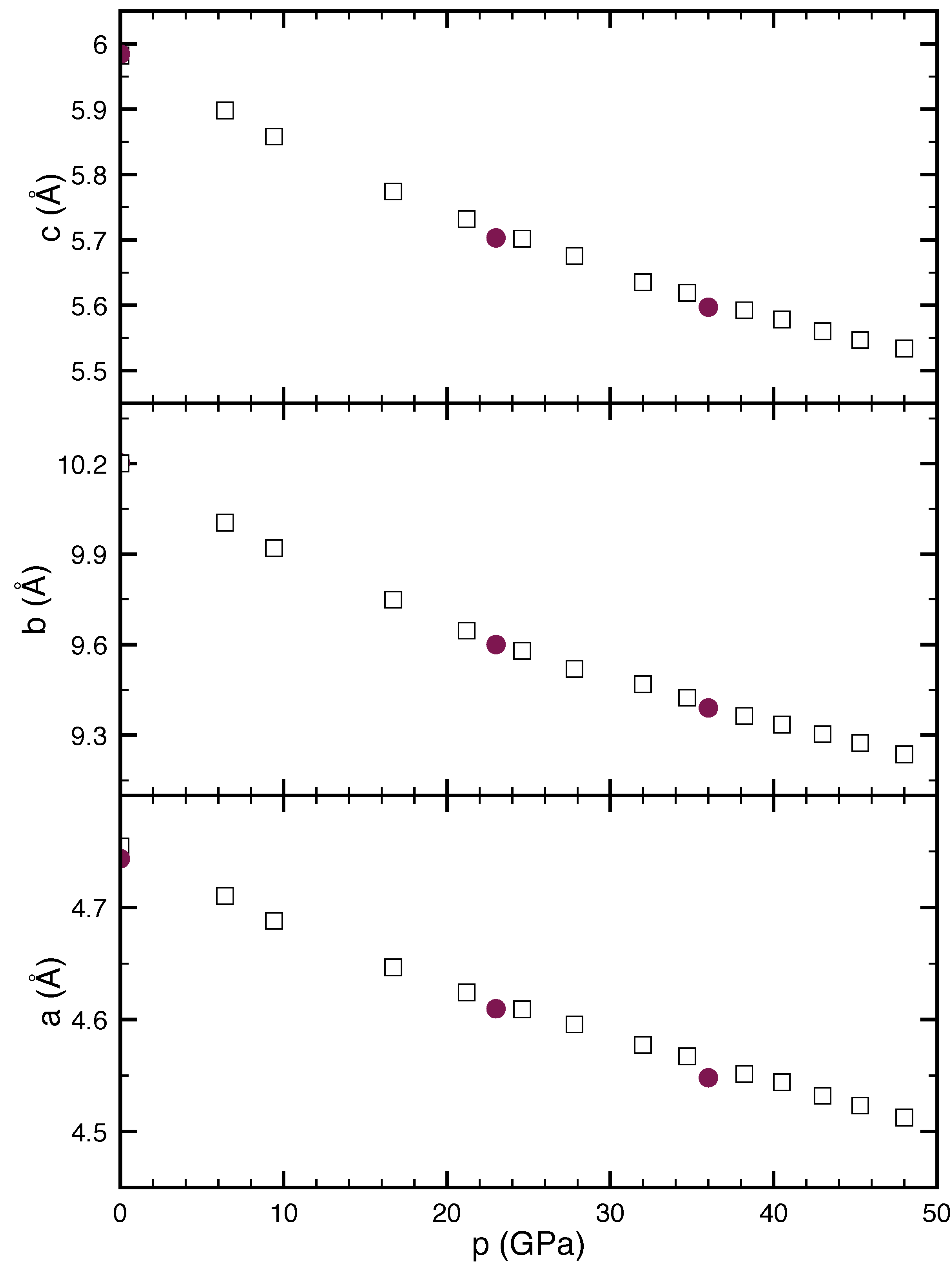}
\includegraphics[width=1.0\linewidth]{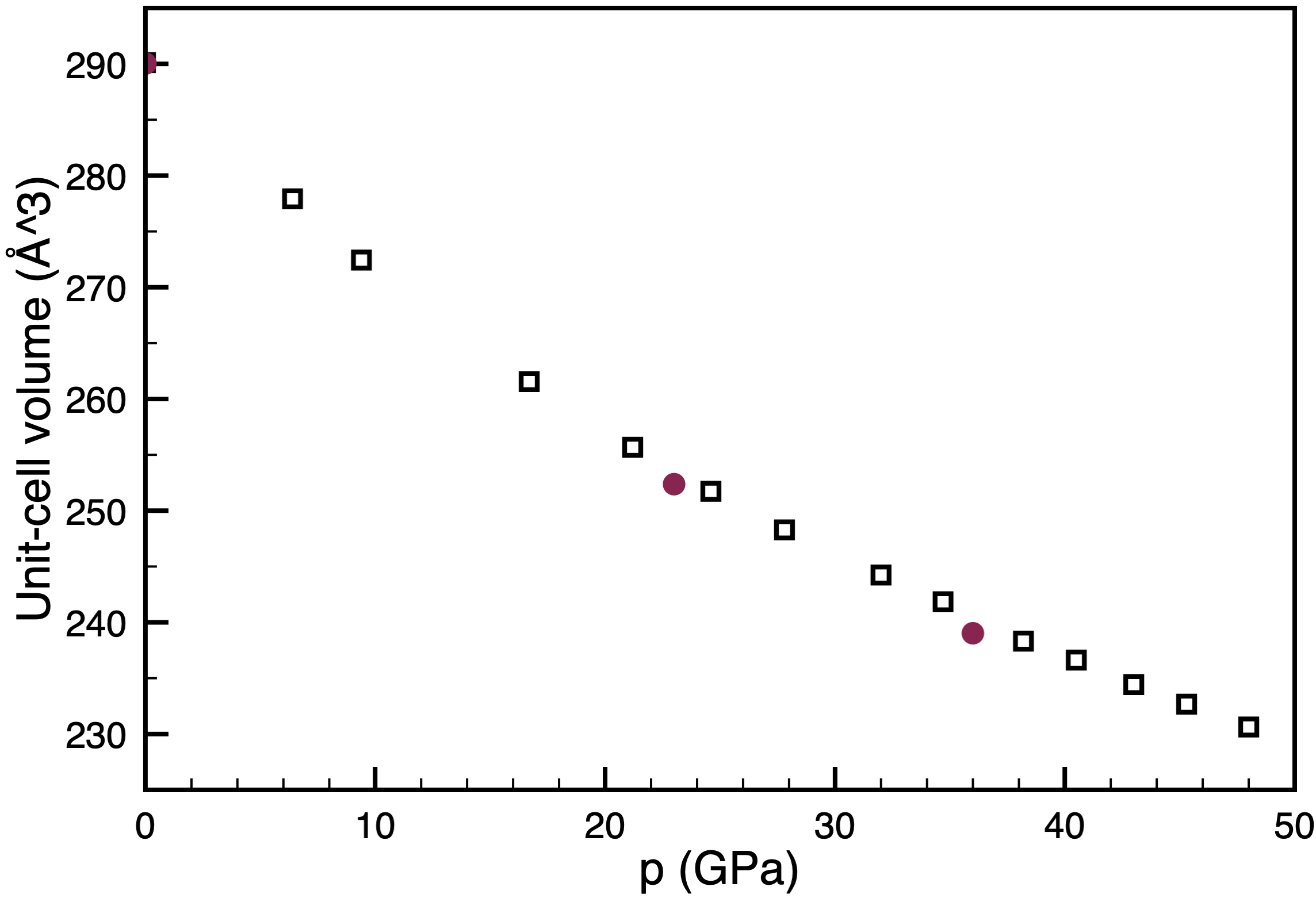}
\caption{The unit-cell parameters of forsterite I measured in this work are plotted along with literature values as a function of pressure. Open squares: Finkelstein et al. (2014) \cite{Finkelstein:2014}, red circles: this work. Error bars are smaller than symbols.
}
\label{FoIabc}
\end{figure}

\begin{table}[h]
\centering
\includegraphics[width=1.0\linewidth]{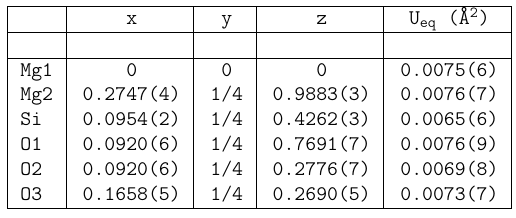}
\caption{
Refined atomic parameters of forsterite I at 23 GPa, 300 K. N$_{all}$= 152, N$_4\sigma$=142, R$_{eq}$=2.3\%, R$_\sigma$=1.9\%, Rall=4.18, R$_4\sigma$=3.94\%.
}
\label{tab}
\end{table}

\begin{figure}
\centering
\subsection*{Forsterite III}
\includegraphics[width=1.0\linewidth]{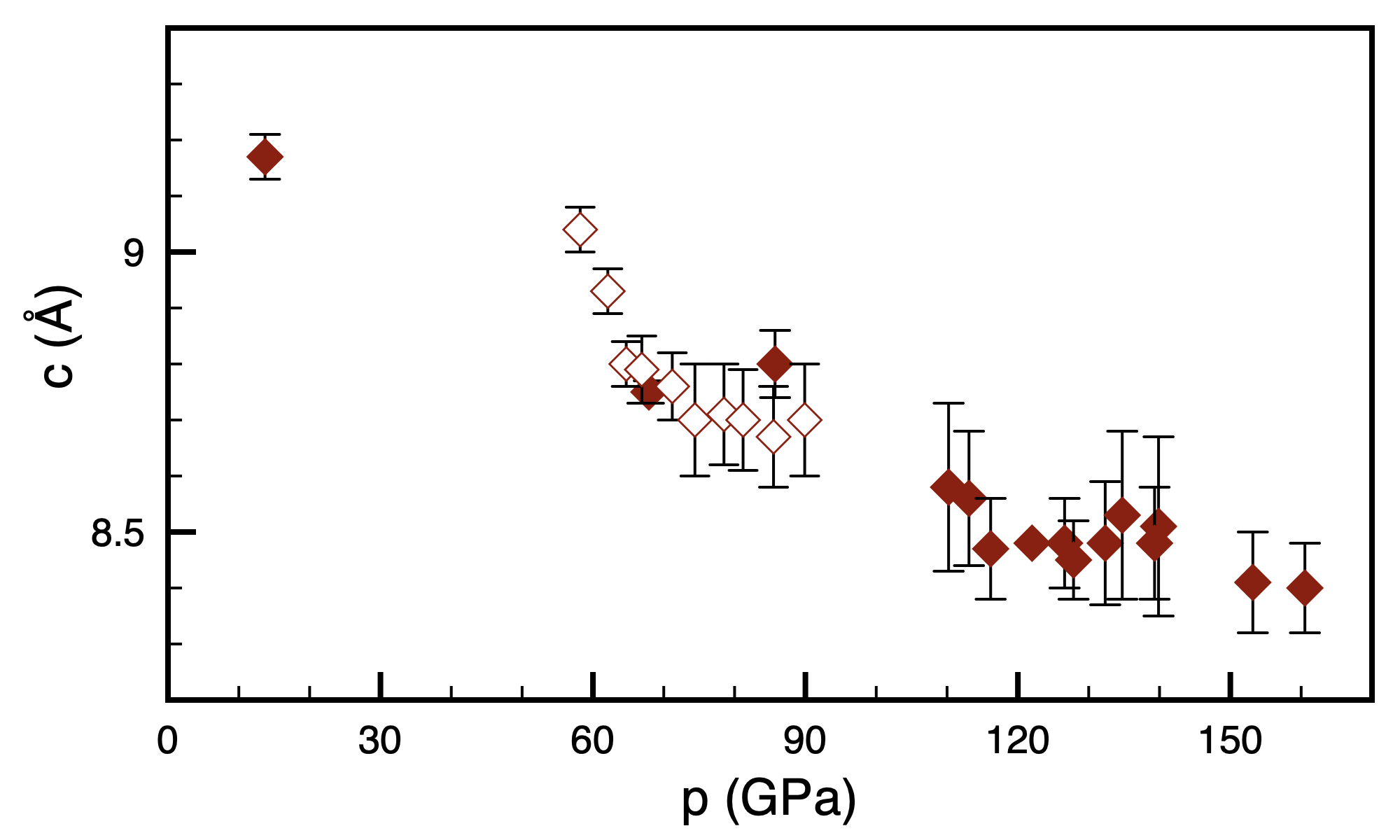}
\includegraphics[width=1.0\linewidth]{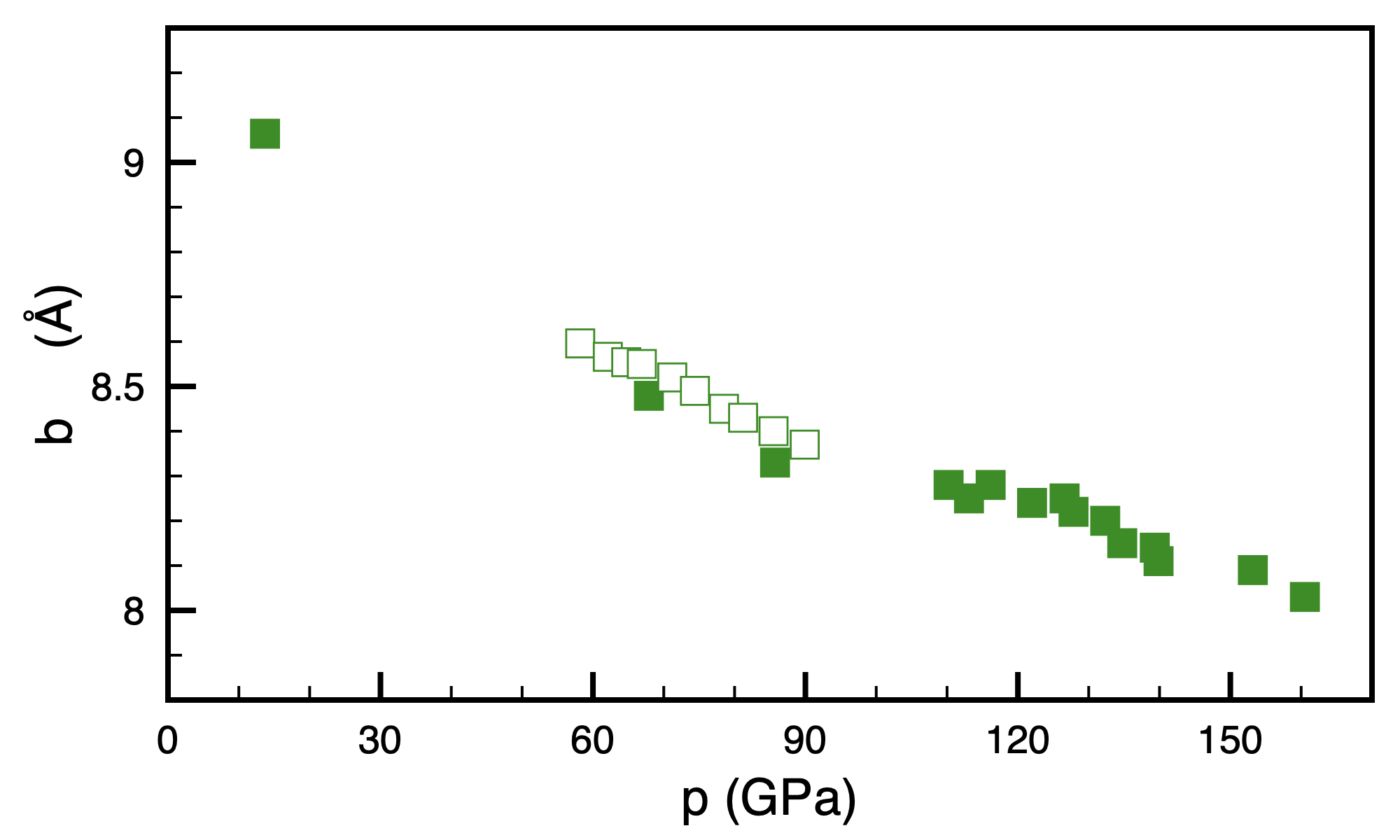}
\includegraphics[width=1.0\linewidth]{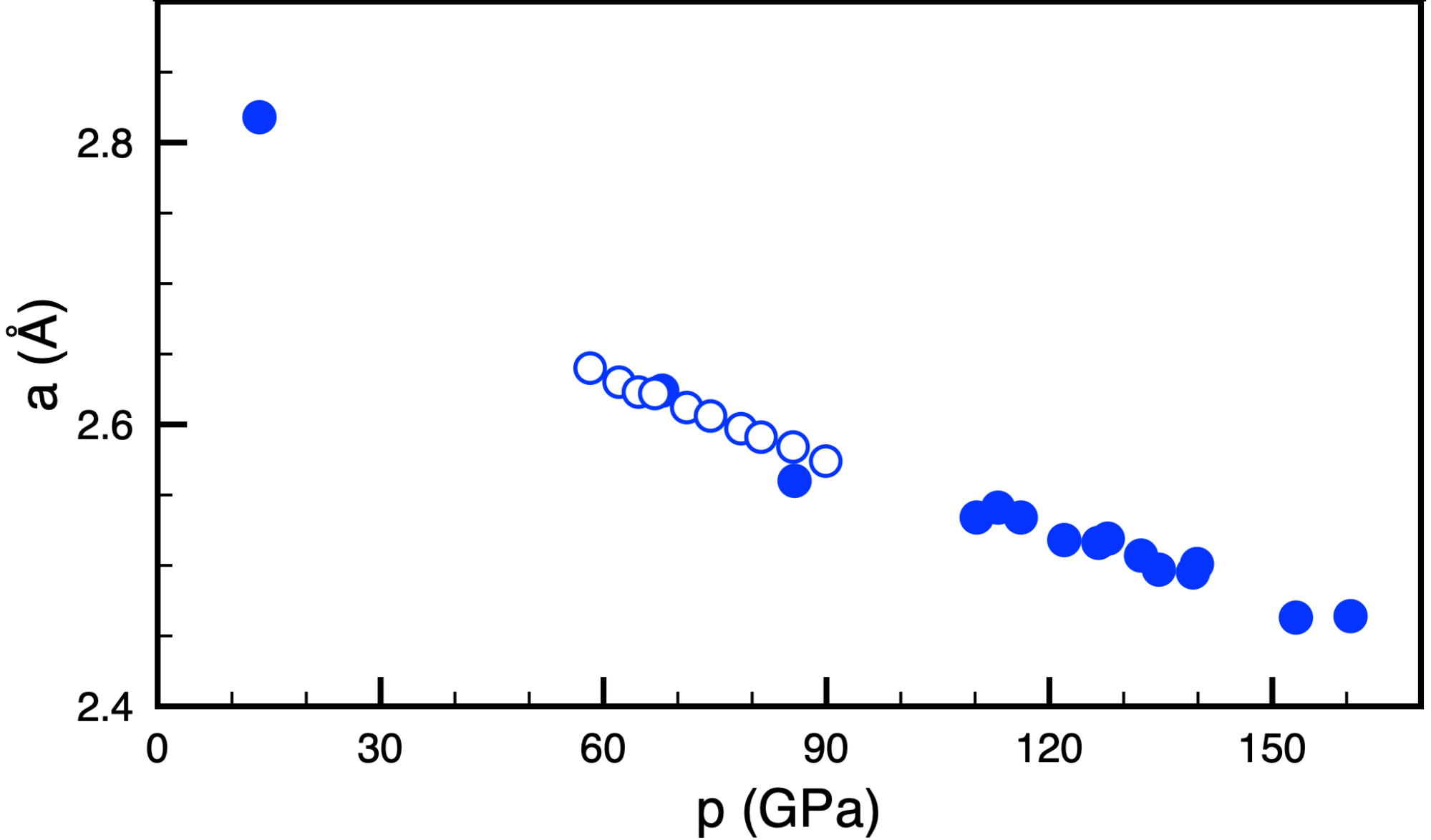}
\caption{
Lattice parameters of forsterite III as a function of pressure. Error bars are smaller than symbols unless shown.}
\label{FoIIIabc}
\end{figure}

\begin{figure*}[th]
\centering
\subsection*{Forsterite III}
\includegraphics[width=1.0\linewidth]{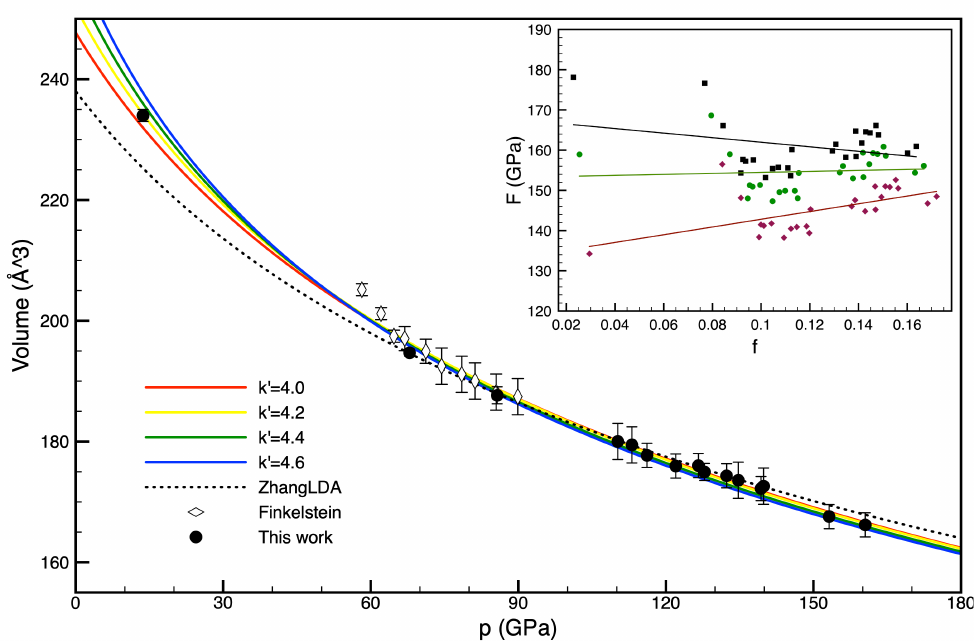}
\caption{
EoS of forsterite III at 300 K. Circles show data measured in this work, diamonds show results of a previous study \cite{Finkelstein:2014}. Black dotted line: calculated EoS from the literature \cite{Zhang:2019aa}; color lines show fits of the $3^{rd}$ order Birch-Murnaham EoS with $K_{0}^{\prime}$  fixed at 4.0 (red), 4.2 (yellow), 4.4 (green), 4.6 (blue). The two lowermost pressure data from Finkelstain et al. \cite{Finkelstein:2014} measured at 58.2 and 62.1 GPa appear to lie off of any of the curves. This is due to the  corresponding \emph{c}-axis off-lying values (Figure \ref{FoIIIabc}), parameter acknowledged by the authors to be the least accurately measured due to diffuse scattering. 
The challenges of fitting an EoS with so few low-pressure information is emphasized in the plot of normalized pressure (F) as a function of Eulerian strain (f) shown in the inset. Data in black, green and red correspond to $V_{0}$ values of 254, 256 and 259 \AA $^{3}$ respectively. The latter shows a better linearity for all observations except for the two lowermost pressure data from Finkelstain et al. \cite{Finkelstein:2014} and a trend consistent with a positive $K_{0}^{\prime}$.
}
\label{EoS}
\end{figure*}

\end{document}

%% file: structure.tex
\usepackage[english]{babel} 

\usepackage{microtype} 

\usepackage{amsmath,amsfonts,amsthm} 

\usepackage[svgnames]{xcolor} 

\usepackage[hang, small, labelfont=bf, up, textfont=sl]{caption} 

\usepackage{booktabs} 

\usepackage{lastpage} 

\usepackage{graphicx} 

\usepackage{enumitem} 
\setlist{noitemsep} 

\usepackage{sectsty} 
\allsectionsfont{\usefont{OT1}{phv}{b}{n}} 

\usepackage{geometry} 

\usepackage[T1]{fontenc} 
\usepackage[utf8]{inputenc} 

\usepackage{cmbright} 

\usepackage{fancyhdr} 
\fancypagestyle{firstpage}{ 
	\fancyhf{}
}

\newcommand{\authorstyle}[1]{{\large\usefont{OT1}{cmss}{eb}{n}\color{Black}#1}} 
\newcommand{\institution}[1]{{\usefont{OT1}{cmss}{m}{sl}\color{Black}#1}} 

\usepackage{titling} 

\newcommand{\HorRule}{\color{Black}\rule{\linewidth}{0.7pt}}

\pretitle{
	\vspace{-30pt} 
	\HorRule\vspace{10pt} 
	\fontsize{36}{20}\usefont{OT1}{cmss}{b}{n}\selectfont 
	\color{DarkGreen} 
}

\posttitle{\par\vskip 12pt} 

\preauthor{} 

\postauthor{ 
	\vspace{8pt} 
	\par\HorRule 
	\vspace{12pt} 
}

\usepackage{lettrine} 
\usepackage{fix-cm}	

\newcommand{\initial}[1]{ 
	\lettrine[lines=3,findent=4pt,nindent=0pt]{
		\color{DarkGreen}
		{#1}
	}{}%
}

\usepackage{xstring} 

\newcommand{\lettrineabstract}[1]{
	\StrLeft{#1}{1}[\firstletter] 
	\initial{\firstletter}\textbf{\StrGobbleLeft{#1}{1}} 
}

\usepackage[backend=bibtex,style=numeric,natbib=true,sorting=none]{biblatex}

\usepackage[autostyle=true]{csquotes} 